\documentclass[review]{elsarticle}

\usepackage{lineno}
\usepackage[colorlinks=true,linkcolor=blue]{hyperref}

\usepackage{graphicx}
\usepackage{caption}
\usepackage{subcaption}

\usepackage{natbib}

\usepackage{xcolor}

\setcitestyle{round}

\journal{.}

\biboptions{authoryear}

\begin{document}

\begin{frontmatter}

\title{Deep Neural Review Text Interaction for Recommendation Systems}

\author{Parisa Abolfath Beygi Dezfouli}

\author{Saeedeh Momtazi\corref{cor2}}
\ead{momtazi@aut.ac.ir}

\author{Mehdi Dehghan}

\address{Computer Engineering and Information Technology Department \\Amirkabir University of Technology, Tehran, Iran}

\cortext[cor2]{Principal corresponding author}

%
%
%

\begin{abstract}
Users' reviews contain valuable information which are not taken into account in most recommender systems. According to the latest studies in this field, using review texts could not only improve the performance of recommendation, but it can also alleviate the impact of data sparsity and help to tackle the cold start problem.
In this paper, we present a neural recommender model which recommends items by leveraging user reviews. In order to predict user rating for each item, our proposed model, named $MatchPyramid\ Recommender\ System$ (MPRS), represents each user and item with their corresponding review texts. Thus, the problem of recommendation is viewed as a text matching problem such that the matching score obtained from matching user and item texts could be considered as a good representative of their joint extent of similarity. To solve the text matching problem, inspired by MatchPyramid \citep{pang_2016}, we employed an interaction-based approach according to which a matching matrix is constructed given a pair of input texts. The matching matrix, which has the property of hierarchical matching patterns, is then fed into a Convolutional Neural Network (CNN) to compute the matching score for the given user-item pair. Our experiments on the small data categories of Amazon review dataset show that our proposed model gains  from 1.76\% to 21.72\% relative improvement compared to DeepCoNN model \citep{zheng_noroozi_yu_2017}, and from 0.83\% to 3.15\% relative improvement compared to TransNets model \citep{catherine_cohen_2017}. Also, on two large categories, namely \texttt{AZ-CSJ} and \texttt{AZ-Mov}, our model achieves relative improvements of 8.08\% and 7.56\% compared to the DeepCoNN model, and relative improvements of 1.74\% and 0.86\% compared to the TransNets model, respectively.
\end{abstract}


\end{frontmatter}


\section{Introduction}
Recommender systems autonomously gather information on the preference of users for a set of items (e.g., hotels, books, movies, songs and etc.) upon which they proactively predict the preference each user would give to an item. Thus, the problem that recommender systems seek to solve could be viewed as a rating prediction problem. Today, with the prevalence of online services and due to the abundance of choice in quite all such services, recommender systems play an increasingly significant role.  

As deep learning approaches have recently obtained considerably high performance across many different machine learning applications, recent researches on recommender systems have also focused on using deep neural networks which have shown impressive results. In this approach, neural networks are mostly used to construct latent representations of users and items using the content associated with them. Content of a user or an item could include any property of each; e.g., demographic information and product preferences are potential content for users, whereas the content linked to items could include their price or any other attribute with respect to their application in general. To make rating prediction, latent representations are then used as input to a regression or collaborative filtering model. However, unlike the content information that describes only the user or only the item, review text could be considered a valuable common context information, featuring joint user-item interaction. Most neural recommender models such as those proposed by \citet{bansal_belanger_mccallum_2016}, \citet{elkahky_song_he_2015}, \citet{kim_park_oh_lee_yu_2016}, \citet{li_kawale_fu_2015} and \citet{wang_wang_yeung_2015} have focused on the content information of either users or items. On the other hand, only a limited number of neural recommender models such as proposed models by \citet{almahairi_kastner_cho_courville_2015}, \citet{seo_huang_yang_liu_2017}, \citet{zheng_noroozi_yu_2017}, \citet{catherine_cohen_2017}, \citet{li2017neural} and \cite{chen2018neural} use user-provided reviews. These review-based models have shown that utilizing reviews could considerably improve the performance of recommendation compared to traditional approaches such as collaborative filtering techniques. Using the valuable information in reviews employed in such models have particularly proved to alleviate the well-known rating sparsity and cold-start problems.

Most review-based neural recommender models work upon text matching approaches; i.e., they work similar to neural information retrieval models with the text corresponding to user as the query and the text corresponding to item as the document. The task of text matching is the central part of many natural language processing applications, such as question answering \citep{xue2008retrieval} and paraphrase identification \citep{socher2011dynamic}. In general, neural text matching models can be divided into two families: representation-based models and interaction-based models. The state-of-the-art neural recommender models which are based on text matching \citep{seo_huang_yang_liu_2017, zheng_noroozi_yu_2017, catherine_cohen_2017}, use the representation-based learning approach. Although interaction-based models have been successfully used for different applications such as question answering, paraphrase identification, and document retrieval, none of the previous proposed neural recommender systems have used this approach. According to \citet{nie2018empirical}, in general, interaction-based approach performs better than representation-based approach in convolutional models for information retrieval. Because with representation-based models, it is very hard to learn global, informative representation of a user document or an item document which is usually very long; it is not surprising that representing every aspect of a long review document with a single vector is almost impossible. On the other hand, interaction-based models determine local matching signals between user document and item document by considering the interaction of every term of user document with every term of item document. It is important to note that traditional information retrieval models essentially work on the basis of similar local matching signals.

In this paper, we propose to model the problem of rating prediction as a text matching problem using an interaction-based approach. To the best of our knowledge, this is the first research to use an interaction-based text matching approach for recommendation. In this formulation, each user is represented by all reviews she/he has written for different items, denoted as user document, and, similarly, each item is represented by all reviews written for it by different users, denoted as item document. Given the representations of all users and all items, our goal is to find the matching score for each pair of user-item for which we intend to determine whether to recommend the item to the user. Based on this idea, we believe that the matching score would be a good representative of the similarity between user and item which is essentially the main guideline for recommendation. Inspired by MatchPyramid \citep{pang_2016}, we employed a CNN architecture fed by the matching matrix of corresponding reviews for a pair of user-item. Our model, MPRS, achieves a better performance with respect to rating prediction compared to the state-of-the-art deep recommendation systems which benefit from users' reviews too.

The rest of this paper is organized as follows: Section 2 describes related works. The proposed model and architecture are introduced in detail in Section 3. The experiments and results are discussed in Section 4, and we conclude the paper in Section 5.

\section{Related Works}
In this section, we provide an overview of the literature that benefits from user reviews for recommendation. We classify this line of research into two categories: non-neural models and neural models.
\subsection{Non-Neural Models}
The \textit{Hidden Factors as Topics} (HFT) model by \citet{mcauley_leskovec_2013} works by regularizing the latent user and item parameters obtained from ratings with hidden topics in reviews. To this end, LDA topic modeling on reviews is combined with a matrix factorization model to be used as an objective function. A modified version of HFT is the $TopicMF$ model by \citet{bao_fang_zhang_2014}, where latent user and item factors learned using matrix factorization are jointly optimized with the topic modeling of their joint review. To this aim, they utilized the non-negative matrix factorization technique. 
\citet{ling_lyu_king_2014} proposed the \textit{Rating Meets Reviews} (RMR) model which extends the HFT model. In their proposed RMR model ratings are sampled from a Gaussian mixture. 

\subsection{Neural Models}
\citet{almahairi_kastner_cho_courville_2015} used matrix factorization to learn the latent factors of users and items. This model benefits from review texts to overcome the data sparsity problem in matrix factorization technique. Following the multitask learning framework by \citet{caruana_1997}, the model jointly predicts rating given by the user $u$ to the item $i$ and models the review written by the user $u$ for the item $i$. The model consists of two components: (1) rating prediction, and (2) review modeling which shares some of the parameters from the former component. \citet{almahairi_kastner_cho_courville_2015} then proposed two representations for modeling the likelihood of the review texts, namely bag-of-words and LSTM embedding. 
\citet{seo_huang_yang_liu_2017} proposed attention-based CNNs to extract latent representations of users and items according to which the model makes rating prediction. One of the recent neural models which efficiently predicts rating is the $Deep\ Cooperative\ Neural\ Networks$ (DeepCoNN) model \citep{zheng_noroozi_yu_2017}. This model consists of two deep neural networks to obtain the latent representations of users and items from their corresponding reviews. User and item representations are then input to a layer to estimate their corresponding rating.  A more recent study which successfully outperformed DeepCoNN on rating prediction is the $Transformational\ Neural\ Networks$ (TransNets) model by \citet{catherine_cohen_2017}. This model is an extension to DeepCoNN model by adding a latent layer to obtain an approximate representation of the review corresponding to the user-item pair in the input. While training, this layer is regularized to be similar to the latent representation of the actual review of the target user-item pair which is computed through training a sentiment analysis network. The main idea is that the joint review of a user-item pair gives an insight into the user's experience with the item. At test time, the joint review is not given; therefore, an approximation of the user-item joint review could improve rating prediction. \citet{li2017neural} proposed a neural recommendation system, named $Neural\ Rating\ and\ Tips\ Generation$ (NRT), which jointly predicts ratings and generates abstractive tips using a multi-task learning approach. \citet{chen2018neural} introduced a neural attention mechanism to simultaneously learn the usefulness of each review and predict ratings. In this setting, highly-useful reviews are obtained to provide review-level explanations which help users to make better and faster decisions.

As mentioned, all the models in the literature have built their architecture on the top of representation-based text matching. On the contrary, the current research is pioneer in utilizing an interaction-based text matching model for neural recommendation to benefit from more local text matching signals.

\section{Model Architecture}
The main idea of the proposed model comes from modeling user-item similarity as text matching, since each user or item could be represented by the text of their corresponding reviews. We perform the text matching task by building a matching/interaction matrix as the joint representation of the user-item pair. To compute the matching score between a user text and an item text, we then use an interaction-based matching model on the top of the matching matrix. In this model, we represent the input of text matching as a matching matrix with entries standing as the similarity between words. 


We consider each data entry as a tuple $(j, k, r_{jk}, rev_{jk})$ which denotes a review written by user $j$ for item $k$ with rating $r_{jk}$ and the review text  $rev_{jk}$.

To this aim, first all the reviews written by user \(j\) are concatenated together into a single document, denoted as \(d_{1:n}^j\), consisting of \(n\) words. The same process is applied to get each item's corresponding document, denoted as \(d_{1:m}^k\) for the item \(k\), where \(m\) is the length of the document. 

The input of our proposed model for the pair of user \(j\) and item \(k\) is a matching matrix, \(M^{j,k}\), with each element \(m_{p,q}^{j,k}\) as the similarity between the words \(w_p\) and \(v_q\) denoting the $p$th and $q$th words in \(d_{1:n}^j\) and \(d_{1:m}^k\) respectively. The similarity between two words can be computed by calculating the cosine similarity of their word embeddings to capture the semantic matching between words. Cosine similarity is calculated as follows:

\begin{equation}
{m}^{j,k}_{p,q}=\frac{{\alpha^j_p}^\top\cdot \beta^k_q}{\big\Vert{\alpha^j_p}\big\Vert\big\Vert{\beta^k_q}\big\Vert} 
\end{equation}

where $\alpha^j_p$ and $\beta^k_q$ are the word embeddings corresponding to \(w_p\) and \(v_q\) respectively. The matching matrix constructed in this way represents the joint representation of the user-item pair.

\begin{figure}[!h]
	\begin{center}
		\includegraphics[scale=0.25]{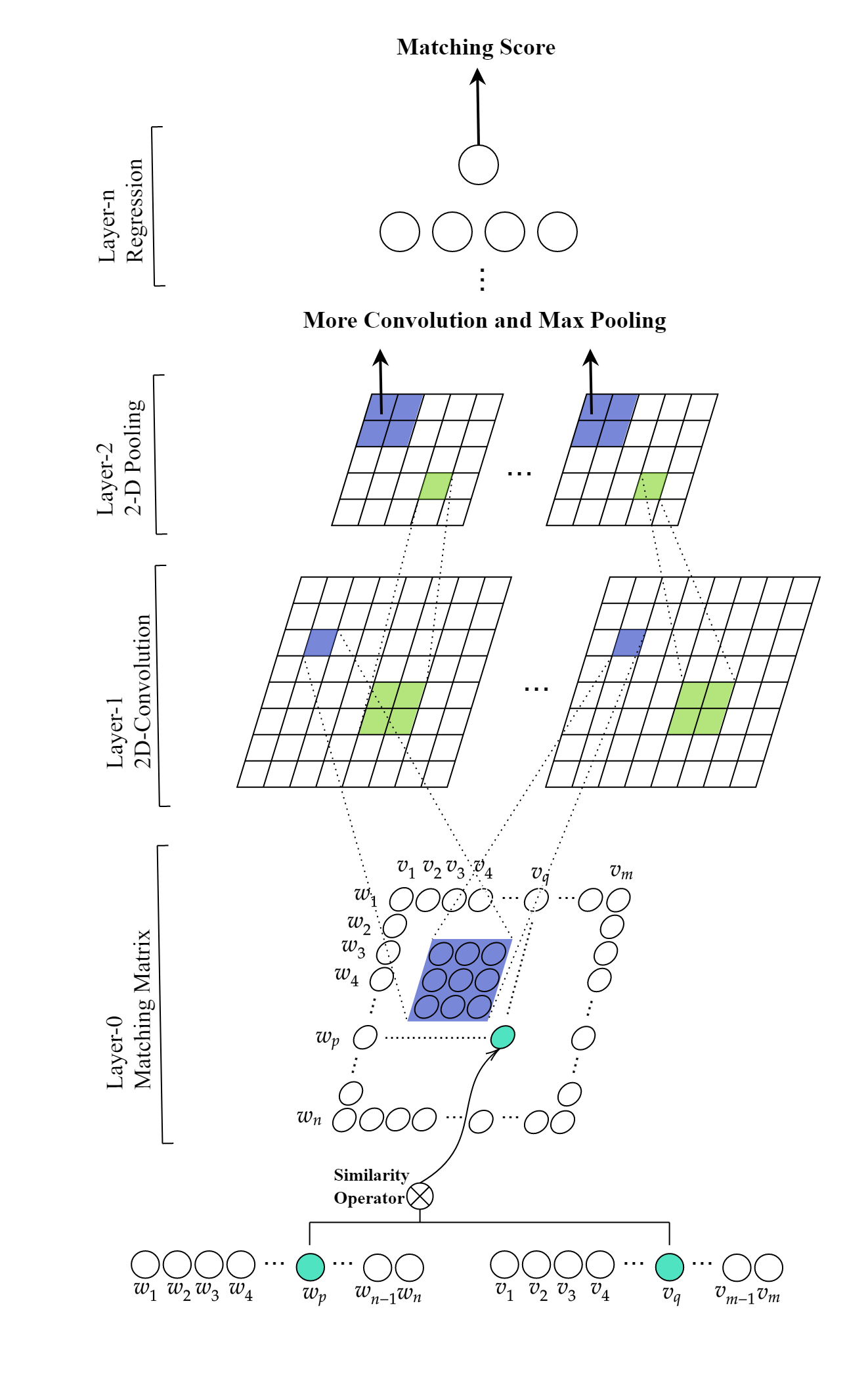}
		\caption{The architecture of the proposed MPRS model}
		\label{fig_1}
	\end{center}
\end{figure}

The architecture of our proposed model for rating prediction is shown in Figure \ref{fig_1}. In the first layer, the matching matrix for the given user-item pair is constructed, as it is described above. 
The matrix, which captures the joint semantic information in the review texts, is then fed into a CNN architecture followed by a regression network to predict the matching score for the corresponding user-item pair.

The CNN part consists of common layers of CNN-based models, including convolution layer, batch normalization layer, and max-pooling layer to learn the underlying interaction patterns. At the end, a fully-connected layer followed by a regression layer is added to the top of the last max-pooling layer to make a prediction on the rating $\hat{r}_{jk}$ that the user $j$ would give to the item $k$.

The CNN model is used to extract effective features in a high-dimensional feature space which has disturbance and distortion. When the effective feature space is obtained, we can use fully-connected layers to perform the desired task of classification or regression. The key operations to extract effective features, or what is called dimension reduction, are filtering and scaling which are performed by convolution and pooling layers respectively.

Each convolution layer uses filter $K_i \in \Re^{c \times t}$ to produce the feature map $z_i$ which is defined as:
\begin{equation}
z_i = f(M^{j,k}*K_i+b_i)
\end{equation}
where $M^{j,k} \in \Re^{n \times m}$ is the input matching matrix, and symbol $*$ is the convolution operator, $b_i$ is a bias term and
$f$ is an activation function. $z_i \in \Re^{(n-c+1) \times (m - t + 1)}$, if the stride is 1. For our model, we use
Rectified Linear Units (ReLUs) as the activation function which is defined as:
\begin{equation}
f(x) = \max({0, x})
\end{equation}

It should be added that we perform a convolution operation regarding each kernel $K_i$ in the convolution layer. Each convolution layer removes infrequent sub-patterns (disturbances) and extracts frequent sub-patterns.

We then apply a max-pooling operation over each feature map obtained from the convolution layer, and take the maximum value in each pooling window as one of the features for the corresponding kernel. In fact, each pooling layer reduces the dimension of the feature maps by sub-sampling. After each pooling operation, 
	the extracted features are more invariant to distortion and scale.

By repeating the convolution and pooling layers, we allow filtering disturbances and concurrently extracting each feature with a certain scale. The result from the final max-pooling layer is passed to a fully connected layer, followed by a regression layer which is a single neuron layer with a linear activation function. Accordingly, the matching score $\hat{r}_{jk}$ is calculated as:
\begin{equation}
\hat{r}_{jk} = W \times O + g
\end{equation}
where $O$ is the results from the fully-connected layer, $W$ is the regression layer's weights and $g$ is the bias.

In essence, the interaction structures in the process of text matching are compositional hierarchies of signals which might appear in the matching matrix, i.e. when matching two texts, word-level matching signals together form the phrase-level signals, and phrase-level signals assemble into sentences-level signals. Thus, the hierarchical convolutions performed in our model would capture the matching patterns between two texts.

The approach is designed based on a deep architecture for text matching, MatchPyramid \citep{pang_2016}. The original architecture was proposed to predict if two texts are similar or not. In our model, however, instead of classifying two texts as relevant/similar or non-relevant/dissimilar, we aim at estimating the relevance degree of two texts. Therefore, our architecture is adapted to make regression prediction for the user-item similarity instead of classification.


\section{Experiments}


\begin{table}[t]
	\begin{center}
		\begin{tabular}{l c c c c c c c c}
			\hline
			Dataset & Category & \#Users & \#Items & \#Ratings \& Reviews\\
			\hline
			AZ-CSJ & Clothing, Shoes and Jewelry & 3,116,944 & 1,135,948 & 5,748,920\\ 
			AZ-Mov & Movies and TV & 2,088,428 & 200,915 & 4,607,047\\ 
			AZ-OP & Office Products & 4,905 & 2,420 & 53,258\\ 
			AZ-IV & Amazon Instant Video & 5,130 & 1,685 & 37,126\\
			AZ-Auto & Automotive & 2,928 & 1,835 & 20,473\\
			AZ-PLG & Patio, Lawn and Garden & 1,686 & 962 & 13,272\\
			AZ-MI & Music Instruments & 1,429 & 900 & 10,261\\
			\hline
		\end{tabular}
		\caption{Dataset statistics separated by category}
		\label{tab:tab_1}
	\end{center}
\end{table}

In order to evaluate the effectiveness of our proposed model, we performed different experiments. In this section, the dataset, the setup of the experiments as well as their results are presented.

\subsection{Dataset}
We evaluate the performance of our proposed model by using the most recent release of Amazon dataset\footnote{\href{http://jmcauley.ucsd.edu/data/amazon/}{http://jmcauley.ucsd.edu/data/amazon/}} \citep{mcauley_pandey_leskovec_2015, mcauley_targett_shi_hengel_2015} which includes reviews and ratings given by users for products purchased on \texttt{amazon.com}. The dataset contains Amazon product reviews and metadata from May 1996 to July 2014, separated by category.
In the following, we present the results of experiments on 7 different categories of this dataset. The statistics of these categories are given in Table \ref{tab:tab_1}. 

The categories presented in the table are from both large and small Amazon datasets. As can be seen, the average number of review and rating pairs provided by each user on the large datasets is less than three, which shows that these categories are extremely sparse. This issue may considerably deteriorate the performance of recommender systems.

More statistics about the length of reviews in these datasets are presented in Table \ref{tab:stats}.

\begin{table}
	\begin{center}
		\begin{tabular}{l c c c c c c c c}
			\hline
			Dataset & User Review Average Length & Item Review Average Length\\
			\hline
			AZ-CSJ & 78.77 & 214.4\\ 
			AZ-Mov & 197.87 & 2047.3\\ 
			AZ-OP & 1298.4 & 2630.65\\ 
			AZ-IV & 545.69 & 1659.32\\
			AZ-Auto & 486.69 & 775.99\\
			AZ-PLG & 1015.79 & 1779.53\\
			AZ-MI & 526.52 & 835.4\\
			\hline
		\end{tabular}
		\caption{Review statistics separated by category}
		\label{tab:stats}
	\end{center}
\end{table}


\vspace{5mm}
\subsection{Evaluation Metric}
In our experiments, we employed the Mean Squared Error (MSE) to evaluate the performance of our proposed model. Let $ N $ be the total number of datapoints in the test set. MSE can be computed as follows:

\begin{equation}
MSE=\frac{1}{N}\sum_{i=1}^N(r_i-\hat{r_i})^2
\end{equation}
where $ r_i $ is the $ i $th observed value and $ \hat{r_i} $ is the $ i $th predicted value.

\subsection{Setup of Experiments}
For each category, we divided the dataset into three sets of training, validation and test which are 80\%, 10\% and 10\% of the whole dataset, respectively. Given training, validation and test sets, the document associated with each user or item is then constructed. Let $ rev_{jk} $ denote the review written by user $ j $ for item $ k $. Let \(d^j\) and \(d^k\) be the documents to be constructed corresponding to $ user_{j} $ and $ item_{k} $ respectively. For all $ (user_{j}, item_{k}) $ pairs in the test set, $ rev_{jk} $ is omitted from both $ d^{j} $ and $ d^{k} $, since at test time, as the simulation of a real world situation, the joint review of a user-item pair is not obviously given. We build the train, validation and test sets accordingly for all our experiments for all the models including the baselines.

\subsection{Baselines}
In order to evaluate the performance of our model, we selected DeepCoNN \citep{zheng_noroozi_yu_2017} and TransNets \citep{catherine_cohen_2017} to compare our results with. They are two of the most recent neural recommender models known as state-of-the-art models in this field. The performance of the DeepCoNN model reported in work by \citet{zheng_noroozi_yu_2017} corresponds to an obsolete version of the dataset to which we had no access, therefore, we implemented the DeepCoNN model and tested the model on the latest version of the Amazon dataset. For the TransNets model and its variant, TransNet-Ext, we used the provided code by the authors that is available on \texttt{github}\footnote{\href{https://github.com/rosecatherinek/TransNets}{https://github.com/rosecatherinek/TransNets}}.

\subsection{Results}
The MSE values of MPRS, DeepCoNN, TransNets and TransNet-Ext models on 7 different categories of Amazon datasets are given in Table \ref{tab:tab_2}. 
\vspace{3mm}

 \begin{table}[!h]
	\begin{center}
		\begin{tabular}{l c c c c}
			\hline
			Dataset & DeepCoNN & TransNets & TransNet-Ext & MPRS (relative improvement)\\ 
			\hline
			AZ-CSJ & 1.5487 & 1.4487 & 1.4780 & 1.4235 (1.74\%)**\\
			AZ-Mov & 1.3611 & 1.3599 & 1.2691 & 1.2582 (0.86\%)*\\
			AZ-OP & 0.7566 & 0.8463	& 0.7495 & 0.7433 (0.83\%)*\\
			AZ-IV & 1.1052 & 1.0564	& 1.0282 & 1.003 (2.45\%)**\\
			AZ-Auto & 1.1758 & 0.9735 & 0.9425 & 0.9204 (2.34\%)**\\
			AZ-PLG & 1.3136	& 1.0958 & 1.0852 & 1.0572 (2.58\%)**\\
			AZ-MI & 1.0705 & 0.9185	& 0.9152 & 0.8864 (3.15\%)**\\
			\hline
		\end{tabular}
		\caption{Performance of the proposed model compared to the state-of-the-art baselines using MSE. * marks statistical significant difference between the proposed model and the the best result from the state-of-the-art baselines at $p < 0.05$ based on 2-tailed paired t-test and ** shows highly statistical significance at $p < 0.01$ based on 2-tailed paired t-test.}
		\label{tab:tab_2}
	\end{center}
\end{table}



As it is shown in the table, our proposed model predict ratings better than the competitive baselines on all the categories. The numbers in the parentheses show the relative improvement compared to the best result in the baseline models. 
As can be seen, our model gains the maximum relative improvement (3.15\%) on \texttt{AZ-MI} category which is the smallest data category in terms of the total number of reviews. It indicates that our model can achieve superior results on small number of instances. Overall, the relative improvements of our model on two large datasets are 0.86\% and 1.74\%, and the relative improvement on small datasets is from 0.83\% to 3.15\% . 

In the next step of our experiments, we evaluate the robustness of our proposed model in case of changing the order of reviews in user/item documents.
According to the structure of the problem and the way we represent users and items in this work, it is expected that if we change the order of reviews in the user document or item document, the model's prediction of rating does not undergo considerable changes. On the other hand, by changing the order of reviews in a document the matching matrix changes as well. As a result, the input to our model would be a different matrix and the following question is raised: ``given our model trained with a specific order of reviews in each user/item document, if we change the order of reviews in user document and item document in test set, does the model's performance in rating prediction undergo considerable changes?''.  

\begin{figure}[h]
	\begin{center}
		\makebox[\textwidth]{\includegraphics[width=80mm,scale=1.5]{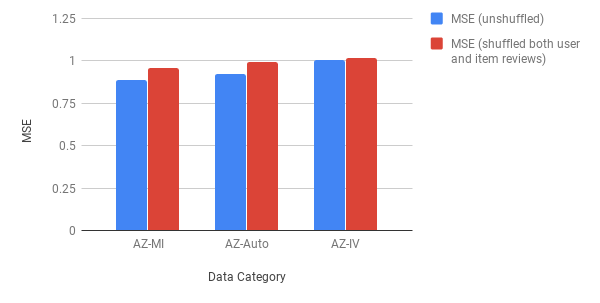}}
		\caption{The impact of shuffling reviews in user and item documents on MSE}
		\label{fig_2}
	\end{center}
\end{figure}

Figure \ref{fig_2} shows the MSE values of MPRS model on three different categories of Amazon dataset. For each category, there are two MSE values reported, one corresponds to the MSE of MPRS on test set with the same order reported in Table \ref{tab:tab_2} and the other with shuffled document of each user and item. 
As can be seen in Figure \ref{fig_2}, for all three categories the amount of changes in the MSE value as a result of changing the order of reviews in documents is insignificant.

\begin{figure}[!h]
	\begin{center}
		\begin{minipage}{.5\textwidth}
			
			\includegraphics[width=1\linewidth]{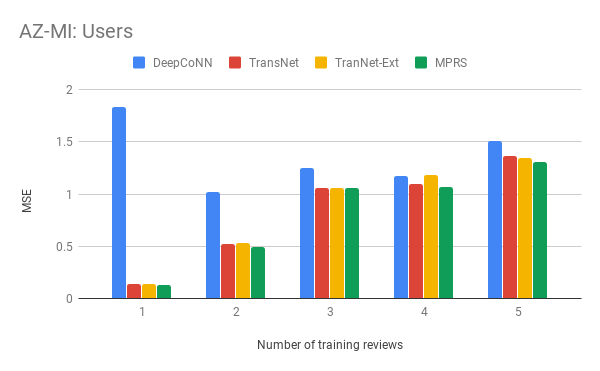}
		\end{minipage}%
		\begin{minipage}{.5\textwidth}
			
			\includegraphics[width=1\linewidth]{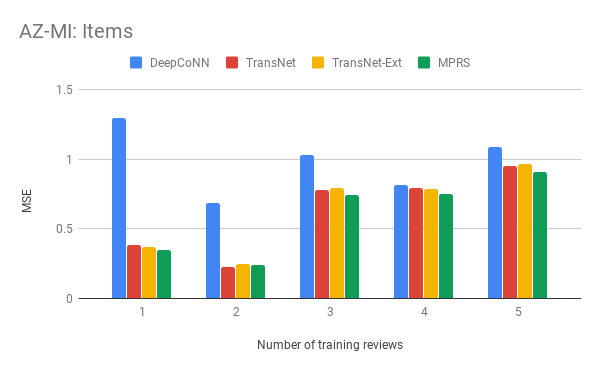}
		\end{minipage}
		\vspace*{-5mm}
		\caption*{(a)}

		\begin{minipage}{.5\textwidth}
			
			\includegraphics[width=1\linewidth]{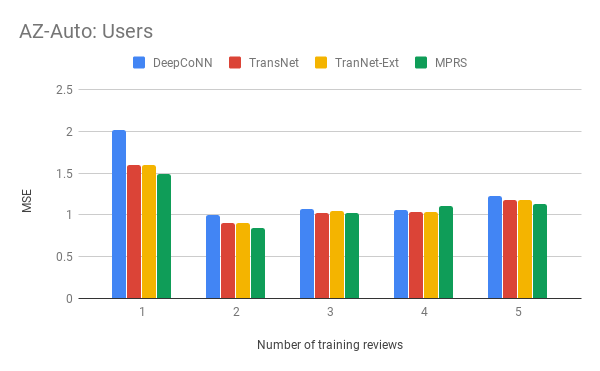}
		\end{minipage}%
		\begin{minipage}{.5\textwidth}
			
			\includegraphics[width=1\linewidth]{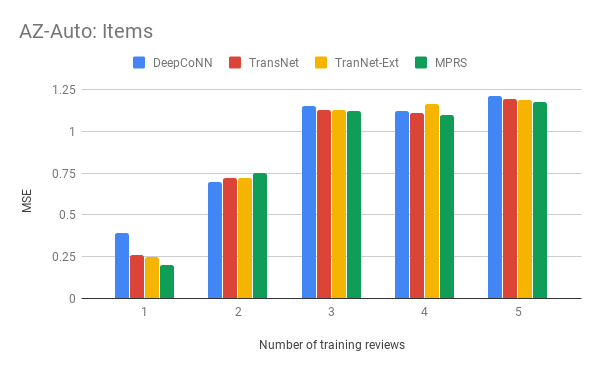}
		\end{minipage}
		\vspace*{-5mm}
		\caption*{(b)}

		\begin{minipage}{.5\textwidth}
			
			\includegraphics[width=1\linewidth]{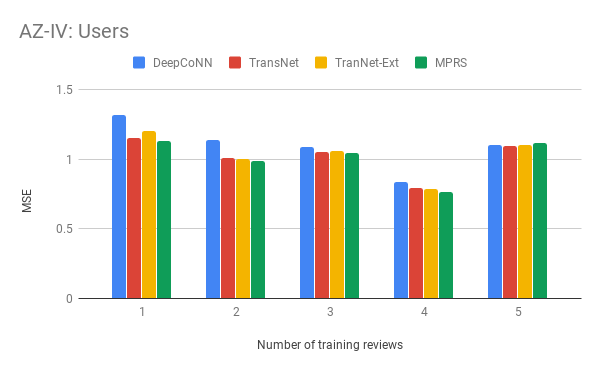}
		\end{minipage}%
		\begin{minipage}{.5\textwidth}
			
			\includegraphics[width=1\linewidth]{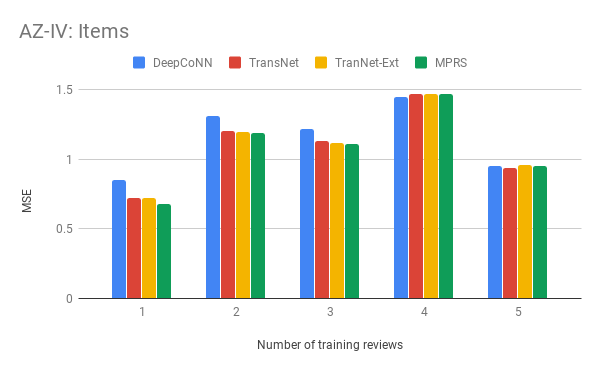}
		\end{minipage}
		\vspace*{-5mm}
		\caption*{(c)}
		
		\caption{MSE values of MPRS compared to DeepCoNN, TransNets and TransNet-Ext for users and items groups with different number of training reviews on three data categories}
		\label{fig_3}
	\end{center}	
\end{figure}

\vspace{3mm}

In the next part of our experiments, we compare the performance of our proposed model with the baseline models in case of data sparsity.
Some users might rate only a limited number of items. Data sparsity problem arises when there is not enough rating data available for some users and items, therefore recommender systems are not able to learn preferences regarding such users and items. Cold start problem, specifically refers to the difficulty of recommendation to new users and items due to lack of data. Data sparsity problem is one of the main challenges in designing recommender systems. Recent studies have shown that using reviews can help to considerably resolve the data sparsity problem. 
The diagrams in Figure \ref{fig_3} depict the MSE values of MPRS, DeepCoNN , TransNets and TransNet-Ext models on three categories of the Amazon dataset. To test the models' performances in this condition, for each dataset category, users and items are grouped based on the number of reviews in training set. In this experiment, the maximum number of training reviews is set to five. For each data category, MSE values of the models are plotted for both users and items groups.
As it is shown in Figure \ref{fig_3}, our proposed model performs better than the baseline models on all three data categories for almost any number of training reviews (1-5).  
Especially, when having only one review available, which is the worst case of data sparsity in this formulation, our model significantly outperforms the baselines.

\section{Conclusion}
In this paper, we propose a novel neural recommender system which recommends items by leveraging user reviews. We represent each user and item with their corresponding reviews and view the problem of recommendation as a text matching problem. We used an interaction-based model to adress the text matching problem. To this aim,  a CNN architecture followed by a regression network was employed to predict user-item pair matching score as a representative of user's rating for item. Experimental results on various data categories of Amazon show that our proposed model outperforms the state-of-the-art neural recommender systems which are based on reviews. 

There are multiple possible directions to extend the model proposed in this paper. One way to extend the model is to use an attention-based CNN to locate the attention to most representative parts of the user-item matching matrix which would probably improve the performance of the base model. Another approach to further improve the performance of the model is to use a target network as well as the main CNN which should be followed by a Transform layer as inspired by the TransNets model \citep{catherine_cohen_2017}. The Transform layer transforms user-item joint feature map obtained from the CNN into an approximation of their joint review which would give an insight into the user’s experience with the item during train as well as evaluation. The target network as used in the TransNets model, provides the latent representaion of target joint review to train the main CNN and the Transform layer.

\bibliography{sample_bib}

\end{document}